\documentclass[runningheads,a4paper]{llncs}

\usepackage{amssymb}
\usepackage{amsmath}
\usepackage{graphicx}
\usepackage{subfig}

\begin{document}

\mainmatter  

\title{Reversible Computation in Wireless Communications}

\titlerunning{Reversible Computation in Wireless Communications}

\author{Harun Siljak\inst{1}}
\institute{CONNECT Centre, Trinity College, The University of Dublin, Ireland\\
\email{harun.siljak@tcd.ie}}

\toctitle{Reversible Computation in Wireless Communications}
\tocauthor{-}
\maketitle

\begin{abstract}
This chapter presents the pioneering work in applying reversible computation paradigms to wireless communications. These applications range from developing reversible hardware architectures for underwater acoustic communications to novel distributed optimisation procedures in large radio-frequency antenna arrays based on reversing Petri nets. Throughout the chapter, we discuss the rationale for introducing reversible computation in the domain of wireless communications, exploring the inherently reversible properties of communication channels and systems formed by devices in a wireless network.

\end{abstract}

\section{Introduction}

Wireless communication systems come in dierent shapes and sizes: from radio frequency (RF) systems we use in everyday life, to underwater acoustic communications (UAC) used where RF attenuation prevents use of radio communications. These two examples are of interest to this case study, as we explored the potential role of reversible computation in improving modern wireless communications in RF and acoustic domain. 

In the RF context, we examine the concept of distributed massive MIMO (multiple input multiple output) systems. The distributed massive MIMO paradigm will have an increasing relevance in fth generation (5G) wireless systems and post-5G era, as it will allow formerly centralised base stations to operate as a group of hundreds (thousands) of small antennas distributed in space, serving many users by beamforming the signal to them, operating using distributed algorithms hence providing reduced power consumption and reduced computational overhead. Our aim is to explore the application of reversible computation paradigms in such systems to contribute in additional reduction of power consumption, but also to help in fault recovery and meaningful undoing of algorithmic steps in control and optimisation of such systems. 

In the underwater acoustic context, we recognised the wave time reversal scheme as a physical example of reversibility, a physical method waiting for its reversible circuit implementation. The mechanism of wave time reversal is analoguous to reversible computation as we know it, and as such it admits elegant and simple circuit implementation benetting from all reversible computation advantages. With this inherent reversibility in mind, we take the question of wave time reversal in underwater conditions a step further, and ask about realistic models of such systems using reversible computation paradigms, and investigate the options of controlling the environment in which this process is used for communication. 

Communication is inherently reversible: the communication channel changes direction all the time, with the transmitter and the receiver changing roles and transmitting through the same medium. Modulation and demodulation, coding and decodingall these processes aim for information conservation and reversibility. Hence the motivation for this study is clear: can reversible computation help in achieving goals of modern wireless communication: increasing access, decreasing latency and power consumption, minimising information losses? 

In this chapter, we present results on optimisation schemes for massive MIMO based on reversing Petri nets, reversible hardware for wave time reversal, and some preliminary thoughts on our work in progress on modelling and control of wave time reversal in reversible cellular automata, as well as control of these automata in general. 

\section{ Reversing Petri nets and Massive MIMO}

\subsection{The Problem}

In the distributed massive MIMO system described in the previous section, not all antennas need to be active at all times. Selecting a subset of antennas to operate at a particular time instant allows the system to retain advantages of a large antenna array, including interference suppression, spatial multiplexing and diversity \cite{r1} while reducing the number of radio frequency (RF) chains and number of antennas to power \cite{r2}. The computational demand of optimal transmit antenna selection for large antenna arrays \cite{r3} makes it impractical, suggesting the necessity of suboptimal approaches. Traditionally, these approaches were centralised and based on the knowledge of the communication channel between every user and every antenna in the array; one widely used algorithm is the greedy algorithm \cite{r4} which operates iteratively by adding the antenna that increases the sum rate the most when joined with the set of already selected antennas. In decentralised algorithms similar procedures are conducted on much smaller subsets of antennas \cite{r5}, leading to similar results in overall performance. Our approach here is decentralised, and it relies on Reversing Petri nets (RPN) \cite{r6} as the underlying paradigm. As this chapter focuses on applications, the reader interested in details about reversing Petri nets used in this example is advised to see \cite{r7}. The presentation here is based on \cite{r8}. 

The optimisation problem we are solving is downlink (transmit) antenna selection of $N_{TS}$ antennas at the distributed massive MIMO base station with $N_T$ antennas, in presence of $N_R$ single antenna users. We maximise the sum-capacity 

\begin{equation}
\mathcal{C}=\max_{\mathbf{P},\mathbf{H}_c}\log_2\det\left(\mathbf{I}+\rho\frac{N_R}{N_{TS}}\mathbf{H}_c\mathbf{P}\mathbf{H}_c^H\right)
\label{eq1}
\end{equation}

where $\rho$ is the signal to noise ratio (SNR), $\mathbf{I}$ a $N_{T S}\times N_{T S}$ identity matrix, $\mathbf{P}$ a diagonal $N_R \times N_R$ power distribution matrix. $\mathbf{H}_c$ is the $N_{T S} \times N_R$ channel submatrix for a selected subset of antennas from the $N_T \times N_R$ channel matrix $\mathbf{H}$ \cite{r9}.

In the case of receiver antenna selection, addition of any antenna to the set of selected antennas improves the overall sum-capacity, as its equivalent of equation (\ref{eq1}) does not involve scaling by the number of selected antennas (i.e. there is not a power budget to be distributed over antennas in the receive case). This problem is submodular and has a guaranteed (suboptimal) performance bound for the previously described greedy algorithm. Greedy algorithm does not haveperformance bound for the transmitter antenna selection, as the case described by equation (\ref{eq1}) does not fullfil the submodularity condition \cite{r10}; the addition of an antenna to the already selected set of antennas can decrease channel capacity. 

As done in \cite{r10,r5}, we optimise (\ref{eq1}) with two variables, the subset of selected antennas and the optimal power distribution over them succesively: first, $\mathbf{P}$ is fixed to having all diagonal elements equal to $1/N_R$ (total power is equal to $\rho N_R/N_{T S}$), and after the antenna selection $\mathbf{P}$ is optimised by water filling for zero forcing. 

\begin{figure}
    \centering
    \includegraphics[width=0.4\textwidth]{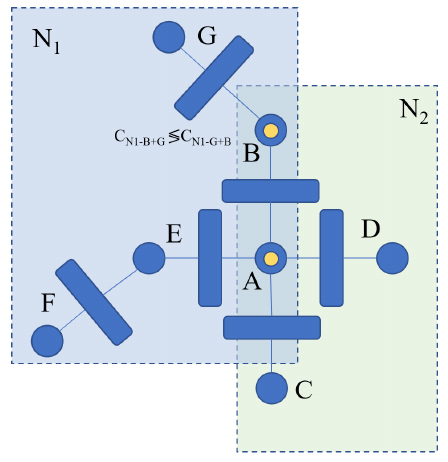}
    \caption{ A toy model of antenna selection on a reversing Petri net}
    \label{f1}
\end{figure}

Fig. \ref{f1} illustrates the proposed algorithm based on RPN: the antennas are Petri net \emph{places} (circles A-G), with the \emph{token} (bright circle) in a place indicates that the current state of the algorithm asks for that place (that antenna) to be on. The places are divided into overlapping \emph{neighbourhoods} ($N_1$ and $N_2$ in our toy example) and each two adjacent places have a common neighbourhood. \emph{Transitions} between places move tokens around based on the sum capacity calculations, with rules described below:

\begin{enumerate}
	\item  Transition is possible if there is a token in exactly one of the two places (e.g. B and G in Fig. \ref{f1}) it connects. Otherwise (e.g. A and B, or E and F) it is not possible.
	
	\item The enabled transition will occur if the sum capacity (\ref{eq1}) calculated for all antennas with a token in the neighbourhood shared by the two places (for B and G, that is neighbourhood $N_1$) is less than the sum capacity calculated for the same neighbourhood, but with the token moved to the empty place (in case of B-G transition, this means
	$\mathcal{C}_{AB} < \mathcal{C}_{AG}$). Otherwise, it does not occur.
	
	\item  In case of several possible transitions from one place (A-E, A-D, A-C) the one with the greatest sum-capacity difference (i.e. improvement) has the
	priority.
	
	\item  There is no designated order in transition execution, and they are performed until a stable state is reached.
	
\end{enumerate}

The algorithm starts from a conguration of $n$ tokens in random places and converges to a stable final conguration in a small number (in our experiments, up to five) of iterations (passes) through the whole network. As the RPN conserves the number of tokens in the network, and our rules allow at most one token per place, the algorithm results in $n$ selected antennas. Executing the algorithm on several RPNs in parallel (in our experiments, up to five) allows tokens to traverse all parts of the network and find good congurations even with a relatively small number of antennas and users. The converged state of the RPN becomes the physical state of antennas: antennas with tokens are turned on for the duration of the coherence interval. At the next update of the channel state information, algorithm proceeds from the current state. 

The computational footprint of the described algorithm is very small: two small matrix multiplications and determinant calculations are performed at a node which contains a token in a small number of iterations. As such, this algorithm is significantly faster and computationally less demanding than the centralised greedy approach which is a low-complexity representative of global optimisation algorithms in antenna selection \cite{r3}. The worst case complexity of RPN based approach is $\mathcal{O}(N_T^{\omega/a})$ (here, $N_T$ denotes the number of antennas, and $\omega$ , $2 < \omega < 3$ is the exponent in the employed matrix multiplication algorithm complexity) if neighbourhood of $N_T^{1/a}$ , $a > 1$ suffices for RPN algorithm (as $\sqrt{N_T}$ suffices in our case, we went for $a = 2$). The constant factor multiplying the complexity is small because of few computing nodes (only those with tokens) and few iterations. 

\begin{figure}
	\centering
	\subfloat[Randomly distributed antennas]
	{
		\includegraphics[width=0.8\textwidth]{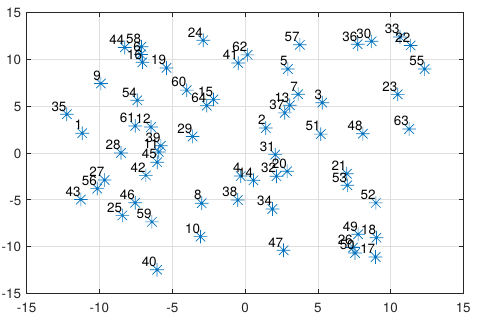}
	}

	\subfloat[The mapping to RPN topology]
	{
		\includegraphics[width=0.8\textwidth]{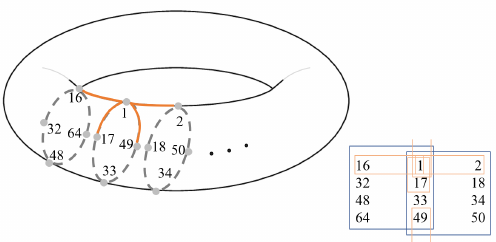}
	}
	\caption{Antennas in physical and computational domain}
	\label{f2}
\end{figure}

\subsection{Results and Discussion}

\begin{figure}
	\centering
	\includegraphics[width=0.8\textwidth]{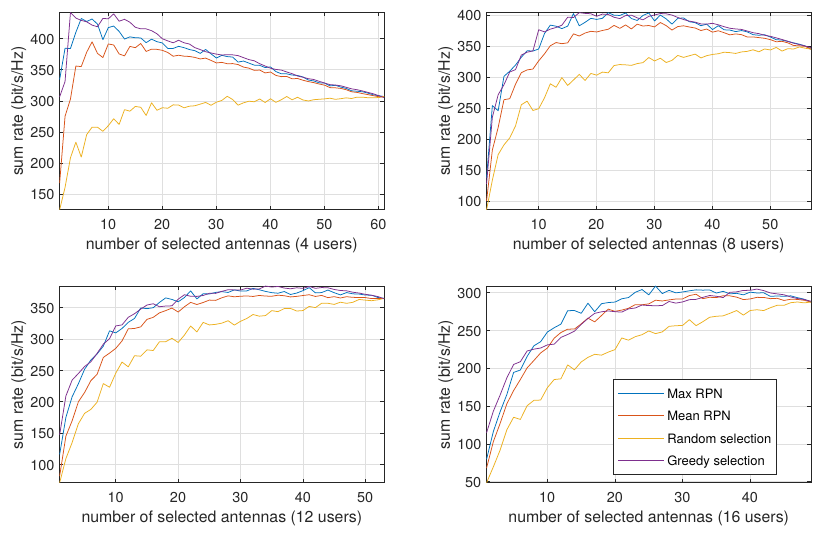}
	\caption{Achieved sum rates for 4-16 users using the proposed algorithm vs random
		and centralised greedy selection}
	\label{f3}
\end{figure}

The algorithm was tested using raytracing Matlab tool Ilmprop \cite{r11} on a system
composed by 64 omnidirectional antennas randomly distributed in space shown in Fig. \ref{f2}(a). In all computations, channel state information (CSI) in matrix $\mathbf{H}$ was normalised to unit average energy over all antennas, users and subcarriers,following the practice from \cite{r9}. 75 randomly distributed scatterers and one large obstacle are placed in the area with the distributed base station. The number of (randomly distributed) users with omnidirectional antennas varied from 4 to 16, and we used 300 OFDM subcarriers, SNR $\rho$ = -5 dB, 2.6 GHz carrier frequency, 20 MHz bandwidth. Antennas are computationally arranged in an
$4 \times 16$ array folded into a toroid, creating a continuous infinite network, as
shown in Fig. \ref{f2}(b), e.g. antenna 1 is direct neighbour of antennas 2, 16, 17 and
49. Immediate Von Neumann (top, down, left, right) neighbours can exchange
tokens, and overlapping 8-antenna neighbourhoods are placed on the grid: e.g. for
antenna 1, transitions to 16 and 17 are decided upon within the neighbourhood
\{16, 32, 48, 64, 1, 17, 33, 49\} and the transitions to 2 and 49 are in \{1, 17, 33, 49, 2, 18, 34, 50\}.
In Fig. \ref{f3} we compare greedy and random selection with two variants of
our RPN approach: the average of five concurrently running RPNs, and the
performance of the best RPN out of those five. The performance is comparable
in all cases, and both variants of our proposed algorithm tend to outperform the
centralised approach as the number of users grows. This in practice means that
a single RPN suffices for networks with a relatively large expected number of
users.

The inherent reversibility of this problem and its solution generalises to the common problem of resource allocation in wireless networks, and sharing any pool of resources (power, frequency, etc) can be handled between antennas (and antenna clusters) over a Reversing Petri Net. At the same time, such a solution would be robust to changes in the environment, potential faults, sudden changes in the mode of operation, and could operate on reversible hardware.

\begin{figure}
	\centering
	\includegraphics[width=0.8\textwidth]{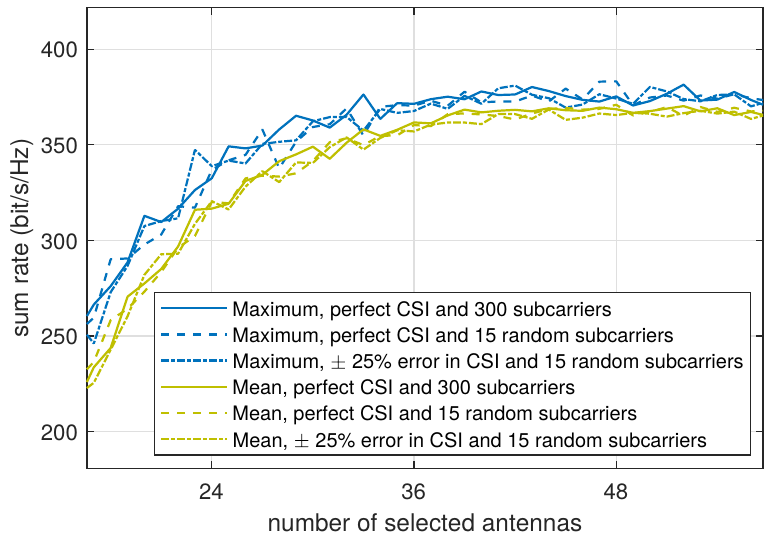}
	\caption{The effects of imperfect CSI and random selection of subcarriers on optimisation}
	\label{f4}
\end{figure}

In \cite{r5}, it has been shown that the distributed algorithms are resistant to errors in CSI and that they perform well even with just a (randomly selected) subset of subcarriers used for optimisation. Results in Fig. \ref{f4} in the case of 12 users confirm this for the RPN algorithm as well.

\section{ Reversible Hardware for Time Reversal}

The technique called wave time reversal \cite{r12} has been introduced in acoustics almost three decades ago, and has since been applied to other waves as well--optical
and RF. In our work, we focused on acoustic time reversal, thinking of its applications in acoustic underwater communications. However, it is worth not-
ing that wave time reversal plays a significant role in RF communications as
well--conjugate beamforming for MIMO systems is based on it. In the remainder
of this section, we introduce the concept of wave time reversal and explain our
proposed solution for its reversible hardware implementation. The presentation
here follows the one in \cite{r13}.

\begin{figure}
	\centering
	\includegraphics[width=0.4\textwidth]{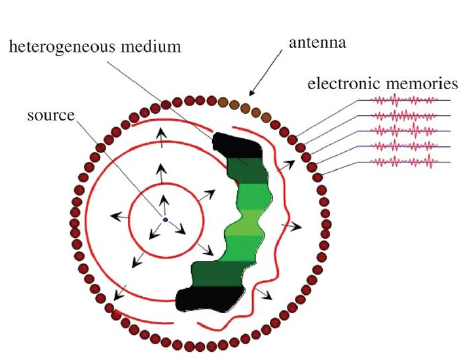}
	\caption{A closed surface is filled with transducer elements. The
		wavefront distorted by heterogeneities comes from a point source and is recorded
		on the cavity elements. The recorded signals are time-
		reversed and re-emitted by the elements. The time-reversed field back-propagates
		and refocuses exactly on the initial source. \cite{r14}
	}
	\label{f5}
\end{figure}

\begin{figure}
	\centering
	\includegraphics[width=0.4\textwidth]{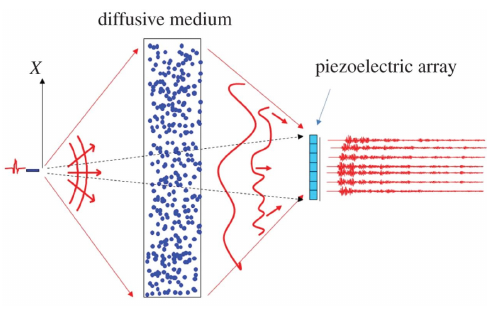}
	\caption{Time-reversal experiment through a diffusive medium \cite{r14}
	}
	\label{f6}
\end{figure}

\subsection{Wave Time Reversal}

Time reversal mirrors (TRMs) \cite{r12} are based on emitter--receptor antennas posi-
tioned on an arbitrary enclosing surface. The wave is recorded, digitized, stored,
time-reversed and rebroadcasted by the same antenna array. If the array on the
boundary intercepts the entire forward wave with a good spatial sampling, it
generates a perfect backward-propagating copy. The procedure begins when the
source radiates a wave inside a volume surrounded by a two-dimensional surface with sensors (microphones) along the surface which record field and its normal
derivative until the field disappears (Fig. \ref{f5}). When this recording is emitted
back, it created the time-reversed field which looks like a convergent wavefield
until it reaches the original source, but from that point it propagates as a diverging wavefield. This can be compensated by an active source at the focusing point canceling the field, or a passive sink as a perfect absorber. \cite{r15}

This description asks for the whole surface to be covered with the TRM transceivers, and for both the signal and the derivative to be stored: for practical
purposes, less hardware-demanding solutions are needed. First, we note that the normal derivative of the field is proportional to the field in case the TRM is in the far field, halving the necessity for signal recording. Second, we note that a
TRM can use complex environments to appear as an antenna wider than it is,
resulting in a refocusing quality that does not depend on the TRM aperture.
\cite{r16} Hence, it can be implemented with just a subset of transceivers located in
one part of the boundary, as seen in Fig. \ref{f6}.

\begin{figure}
	\centering
	\subfloat[]
	{
		\includegraphics[width=0.4\textwidth]{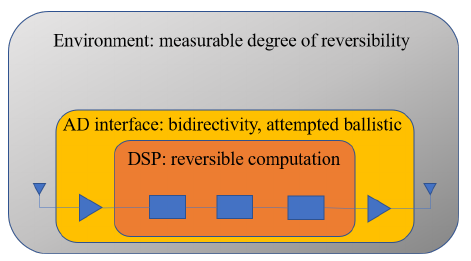}
	}
	\subfloat[]
	{
		\includegraphics[width=0.4\textwidth]{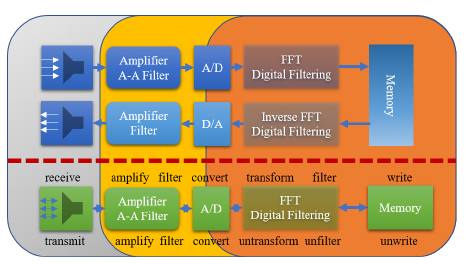}
	}
	\caption{(a) The three realms of reversibility, (b) The classical (top) and the reversible solution (bottom) for the classical time reversal chain
	}
	\label{f7}
\end{figure}

\subsection{The Design}

Fig. \ref{f7} illustrates the challenge of designing reversible hardware solution for a
TRM:

\begin{enumerate}
	\item   The environment is reversible to an extent (we will return to this question
	later in this chapter). The physics of wave propagation in water is reversible,
	but the issues arise as we lose information in the process.
	
	\item  The analog computation part of the TRM loses information due to filtering
	and analog-to-digital/digital-to-analog conversion (ADC/DAC), amplifiers
	accompanying the filters and the converters themselves, at the transition to
	the digital domain.

	\item Finally, the digital computation part of the TRM is reversible and no increase
	in entropy is necessary: writing in memory and unwriting, in the fashion of
	Bennett's trick, enabling reuse of memory for the next incoming wave, while
	not increasing the entropy.

\end{enumerate}

\subsubsection{Analog processing} 

The real amplifier is an imperfect device with a limited
bandwidth, hence prone to losing signal information. By definition, it takes additional energy for the signal, so it asks for an additional power source. At the same
time, the analog to digital and digital to analog converters both lose information because of the finite resolution in time and amplitude, preventing full reversibility.
However, a single device can be both an ADC and a DAC depending on the direction \cite{r17}. In this solution, we assume put bi-directional converters together with bi-directional amplifiers \cite{r17}. The conversion is additionally simplified in
the one-bit solution \cite{r18} where the receivers at the mirror register only the sign
of the waveform and the transmitters emit the reversed version based on this
information. It is a special case of analog-to-digital and digital-to-analog conversion with single bit converters. The reduction in discretisation levels also means
simplification of the processing chain and making its reversal (bi-directivity)
even simpler. The question of the information loss is not straightforward: while
the information about the incoming wave is lost in the conversion process (and
the loss is maximal due to minimal resolution), spatial and temporal resolution
are not significantly degraded. This scheme can also be called "one-trit" reversal:
there are three possible states in the practical implementation: positive pressure,
negative pressure, and "off".

\subsubsection{Digital processing}

 The first, straightforward way of performing time reversal
of a digitally sampled wave is storing it in memory and reading the samples in
the reverse order (last in, first out, LIFO), analoguous to storing the samples
on the stack. The design of registers in reversible logic is a well-explored topic
\cite{r19} and both serial and parallel reading/writing can be implemented. Design
of latches in reversible logic is a well-studied problem with known solutions;
a combination of latches makes a flip-flop, and a series of flip-flops makes a
register (and a reversible address counter). In the case of wave time reversal,
the recording of data is a large register being loaded serially with wave data.
$m$ bits from the ADC are memorised
at the converter's sample rate inside a
$k \times m$ bit register matrix (where $k$ is
the number of samples to be stored for
time reversal). In the receiving process, the bits are stored, in the transmission
process they are
unstored, returning the memory into the blank state it started
from (uncomputation). We utilise Bennett's trick and lose information without
the entropic penalty: the information is kept as long as it is relevant.

When additional signal processing, e.g. filtering or modulation is performed,
it is convenient to reverse waves in frequency domain: there, time domain reversal
is achieved by phase conjugation, i.e. changing the sign of the signal's phase. The
transition from time to frequency domain (and vice versa) in digital domain is
performed by the Fast Fourier Transform (FFT) and its inverse counterpart,
which are reversibly implementable \cite{r20}. The necessary phase conjugation is
an arithmetic operation of sign reversal, again reversible. Any additional signal
processing can be reversible as well: e.g. filter banks and wavelet transforms.
These processes remain reversible with preservation of all components of signals
\cite{r21}.

\begin{figure}
	\centering
	\subfloat[]
	{
		\includegraphics[width=0.8\textwidth]{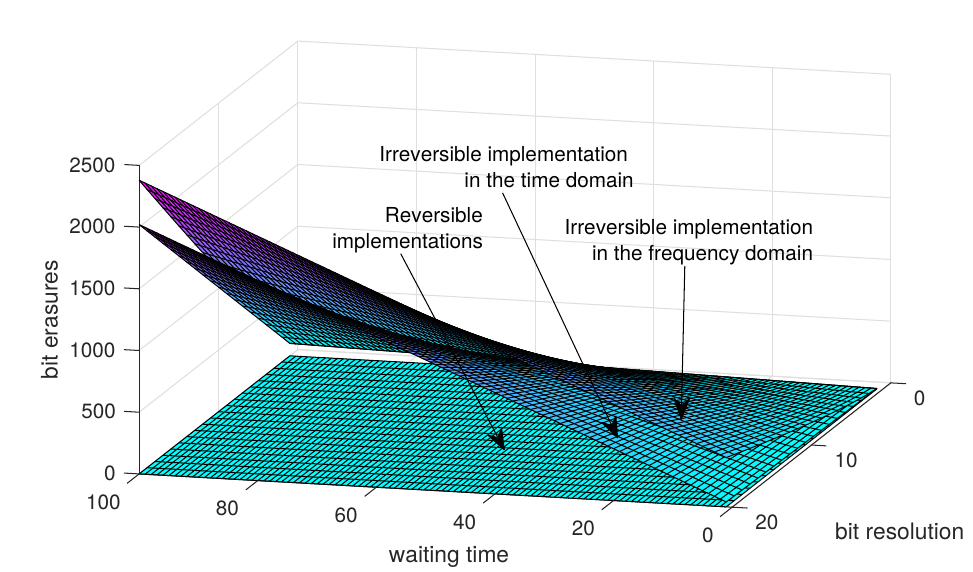}
	}

	\subfloat[]
	{
		\includegraphics[width=0.8\textwidth]{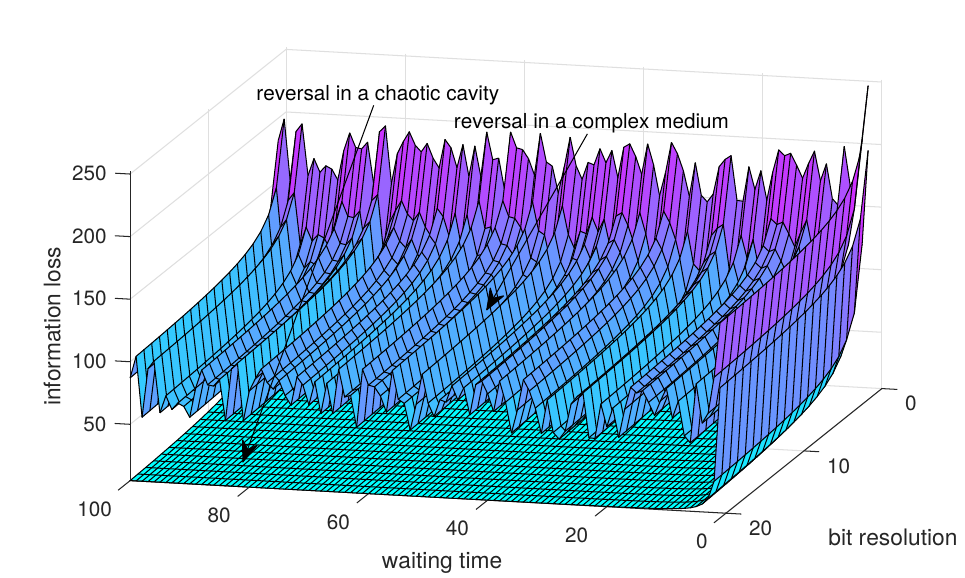}
	}
	\caption{Information loss in (a) digital and (b) analog part of the system. Units
		are omitted as the particular aspects of implementation are not relevant for the
		illustration of effects. Plot (a) is obtained by counting operations, plot (b) by
		simulation of back-scattering.}
	\label{f8}
\end{figure}

Fig. \ref{f8}(a) gives a comparison of the bit erasures in different implementations of the digital circuitry: frequency domain (FFT) and time domain reversal
performed by irreversible circuits, compared to reversible implementations. The
number of erasures changes depending on two parameters: bit resolution of the
ADC and the waiting time--the length of the interval in which samples are collected before reversal starts, equivalent to the number of digitised samples. The
increase in both means additional memory locations and additional dissipation
for irreversible circuits. The irreversible FFT implementation has an additional
information loss caused by additional irreversible circuitry compared to the irre-
versible time domain implementation. Our implementation has no bit erasures
whatsoever. The price that is paid reflects in the larger number of gates used
in the circuit: the number of gates has only spatial consequences, information-
related energy dissipation is zero thanks to information conservation.

On the other hand, Fig. \ref{f8}(b) shows the information loss in the analog part
of the system, and we differentiate two typical environments, the chaotic cavity
and the complex (multiple scattering) medium. The chaotic cavity is an ergodic
space with sensitive dependence on initial conditions for waves. In such an en-
vironment there is little to no loss in the information if the waiting time is long
enough and the ADC resolution is high enough. In the complex media, the difference is caused by some of the wave components being reflected backwards by
the scattering environment, hence not reaching the TRM. Again, more information is retained with the increase in the ADC resolution. However, as reported
in \cite{r18}, the information loss from low-resolution ADC use does not affect the
performance of the algorithm. The analog part of the scheme remains a topic of
our future work, as it leaves space for improvements of the scheme.

\section{Reversible Environment Models and Control}

Time reversal described in the previous section is an example of a reversible
process in a nominally reversible environment. While dynamics of water subject
to waves are inherently reversible, most of the sources of the water dynamics
do not reverse naturally: e.g. the Gulf stream or a motion of a school of fish.
Hence, even though it would rarely be completely reversed, the model for UAC
should be reversible. We discuss the questions of reversible models following
the exposition in \cite{r22}, and the work in progress on control of reversible cellular
automata (RCA).

\begin{figure}
	\centering
	\includegraphics[width=0.5\textwidth]{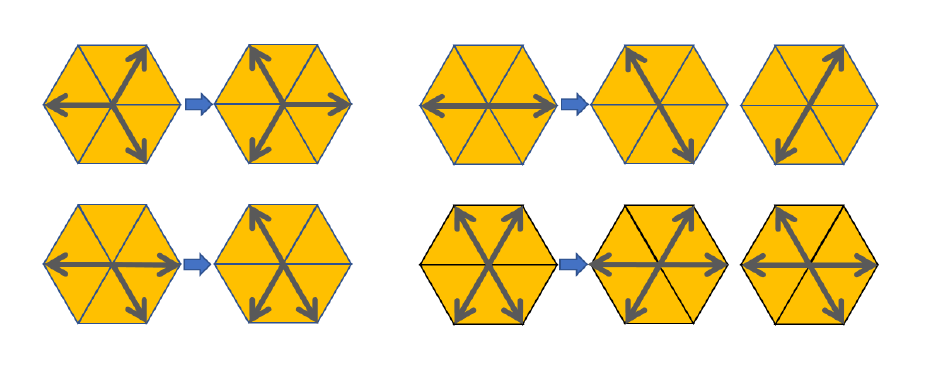}
	\caption{FHP rules }
	\label{f9}
\end{figure}

RCA lattice gas models are cellular automata obeying the laws of fluid dynamics described by the Navier-Stokes equation. One such model, FHP (Frisch-
Hasslacher-Pomeau) lattice gas \cite{r23} is simple and yet following the Navier-Stokes
equations exactly. It is defined on a hexagonal grid with the rules of particle collision shown in Fig. \ref{f9}. The FHP lattice gas provides us a two-dimensional model
for UAC, easily implementable in software and capturing the necessary properties of the reversible medium.

Following the exposition in the previous section, we observe a model with an
original source (transmitter) which causes the spread of an acoustic wave, the original sink (receiver) waiting for the wave to reach it, as well as scatterers and
constant flows (streams) in the environment. The constant stream and the loss
of information caused by some wave components never reaching the sink will
result in an imperfect reversal at the original source. The measure of returned
power gives us a directivity pattern (focal point). The amplitude of the peak
will fluctuate based on the location of the original source and may is a measure
of reversibility, akin to fidelity or Loschmidt Echo. For us, it is a measure of
the quality of communication, but in a more general context it can measure
reversibility of a cellular automaton.

From the control viewpoint, it is interesting to ask the following: if a certain
part of the environment is controllable (i.e. a number of cells of the RCA does
not obey the rules of the RCA but allows external modification), how can it be
used to achieve better time reversal? This is a compensation approach where
we engineer the environment to compensate for effects caused by sources of
disturbance out of our control. The approach we take is one of control of cellular
automata \cite{r24}, and it will be shown that RCA are easier to control than regular
CA, with easier search strategies and the ability to calculate control sequences.

\section{Conclusions}

In this chapter, we provided an overview of results obtained in the case study on
reversible computation in wireless communications. Some of the presented work,
such as optimisation in massive MIMO and reversible hardware for wave time
reversal is finished and subject to further extensions and generalisations; other
work, mainly the parts focused on RCA and modelling of reversible physics of
communication, is still ongoing and more results are to come. This has been a
pioneering study into reversibility in communications, and the results obtained
promise a lot of space for improvement and applications in the future, and we
hope these efforts will serve as an inspiration and a trigger for the development
of this field of research.

\section{Acknowledgements}

The work presented in this chapter was supported by the COST Association
through the IC1405 Action on Reversible Computation, as well as a grant from Science Foundation Ireland (SFI) co-funded under the European Regional Development Fund under Grant Number 13/RC/2077 and European Union's Horizon 2020 programme under the Marie Skaodowska-Curie grant agreement No
713567. I am grateful to my collaborators, Prof Anna Philippou, Kyriaki Psara,
Dr Julien de Rosny, Prof Mathias Fink, and Dr Franco Bagnoli for making this
interdisciplinary research possible, and to K. Popovic for the inspiring ideas.

\end{document}